\documentclass{nature}

\usepackage{graphicx}
\usepackage{color}

\title{Neutrinos from tidal disruption events}

\author{Kimitake Hayasaki$^{1}$}

\begin{document}

\maketitle

\begin{affiliations}
 \item  Department of Astronomy and Space Science, Chungbuk National University, Cheongju 361-763, Republic of Korea
\end{affiliations}

\begin{abstract}
Tidal disruption events are an excellent probe for supermassive black holes in distant inactive galaxies because they show bright multi-wavelength flares lasting several months to years. AT2019dsg presents the first potential association with neutrino emission from such an explosive event. 
\end{abstract}

According to the Big Bang theory, the neutrino is the second most common elementary particles in our universe after photons [1,2]. Neutrinos are called ghost particle because they so weakly interact with matter that it is difficult to detect them. However, this is an advantage on another front: neutrinos {carry} direct physical information about astronomical phenomena so that we can understand them more deeply. High energy astrophysical neutrinos are produced by interacting relativistically accelerated cosmic-rays with an ambient matter or photons. 
While the observation of astrophysical neutrinos has increased in recent years, they are often detected without a clearly identifiable source. Only three astrophysical sources of neutrinos have been identified so far: the Sun, the 1987A supernova, and the blazar TXS~0506+056 [3], but it remains in debate for the association with TXS~0506+056. The first two were detected by Homestake, Kamiokande, and Super-Kamiokande [4], which are sensitive to low-energy neutrinos, while the blazar neutrino was detected by IceCube, which is sensitive to very high-energy neutrinos. In this issue of Nature Astronomy, Robert Stein and collaborators [5] report that a recently detected high-energy IceCube neutrino, {IceCube-191001A}, is associated with the tidal disruption event (TDE) AT2019dsg. This neutrino has an energy of $\sim0.2$ PeV and is thus the second most energetic astrophysical neutrino source, with energies above 100 TeV, after the blazar TX0506+056 [3]. 

A TDE occurs when a star on a Keplerian orbit gets close enough to the supermassive black hole (SMBH)  to be disrupted by the SMBH's tidal forces. Then the stellar debris falls back to the SMBH at a {super-Eddington} rate, showing a characteristic flare that lasts for months to years [6]. TDEs are known as among the brightest transient phenomena in our universe over a wide range of wavebands from optical to X-rays and, therefore, work as the excellent probes of dormant SMBHs at the centers of distant inactive galaxies. Recent multi-wavelength observations have revealed the diverse properties of TDEs [7,8]. TDEs are divided into two categories: thermal TDEs without a relativistic jet and non-thermal TDEs with a relativistic jet (so-called jetted TDEs). Remarkably, most thermal TDEs shine brightly only in soft-X-ray wavebands (soft-X-ray TDEs) or in optical/UV wavebands (optical/UV TDEs). However, AT2019dsg is an unusual type of TDE because it shows bright emission from optical to soft-X-ray wavebands as well as weak but observable radio emission [5,8]. Fig.~1 depicts a hypothetical picture of a disk-outflow-jet system after tidal disruption of a star by an SMBH to explain the observed diversity. The IceCube-191001A-AT2019dsg association can be a key to help us understand the observed diversity of TDEs.

The probability that IceCube-191001A has an astrophysical origin is estimated to be 59\% from a simple energetics argument [5]. The possibility that it is of an atmospheric origin cannot thus be excluded completely. While the IceCube probability is only $59\%$, if the number of atmospheric neutrinos is low, the temporal and spatial association with AT2019dsg increases the probability that the two are associated. In the following, we assume that the neutrino was emitted from AT2019dsg and explore the relevant physical mechanism that could have caused it. According to the blazar neutrino analogy [3], it is natural to regard that the TDE neutrino produced from a relativistic jet. As a companion paper of Stein et al. (2021), Walter Winter \& Cecilia Lunardini (2021) [9] propose a model in which neutrinos are generated from internal shocks in a relativistic jet by a photo-meson interaction. In their model, neutrino production is driven by back-scattered X-ray photons inside the outflow, which are delivered to the plasma shell (shocked region) that travels inside the jet. While the neutrino production rate increases as the density of supplied photons increases, the production efficiency decreases as the size of the plasma shell increases. The balance of these two explains {the} $\sim150$ days delay between the neutrino detection and the observed optical/UV peak of the TDE.

\begin{figure}
\begin{center}
\includegraphics[width=\linewidth]{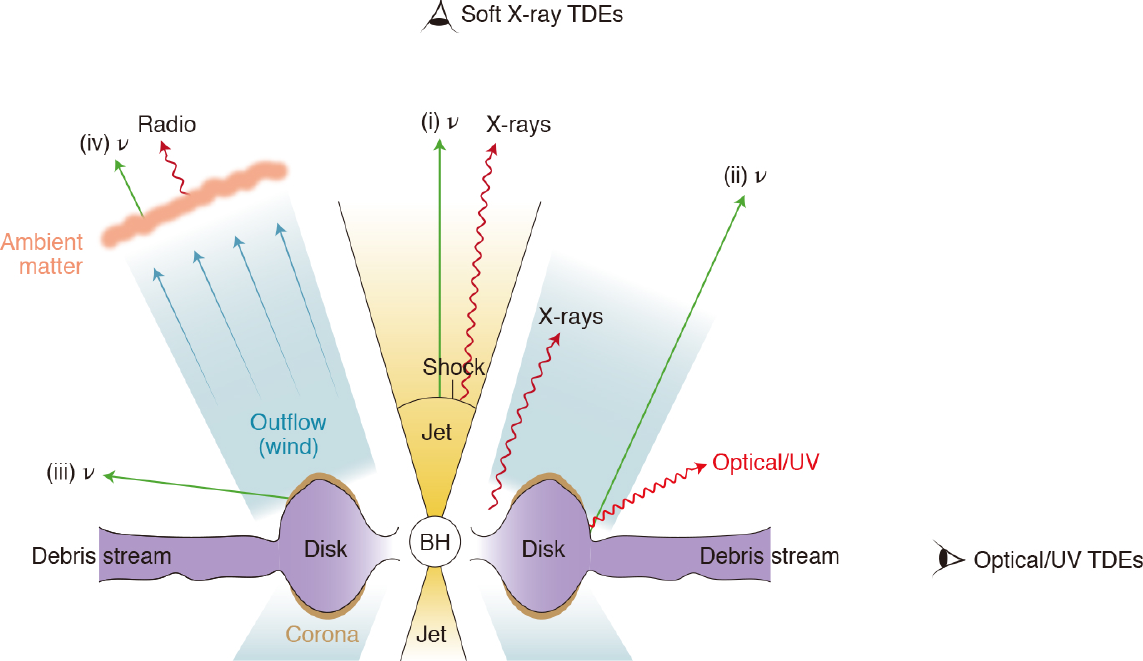}
\end{center}
 \caption{
An illustration of the disk-outflow-jet system formed after the tidal disruption of a star, as in the case of AT2019dsg. Here $\nu$ indicates a neutrino. Depending on the viewing angle, the waveband of observable thermal emission from TDEs changes from soft-X-rays to the optical/UV [15]. AT2019dsg shows optical to X-ray variability with weak radio emission [5,8]. It has been proposed that the high-energy neutrino is produced in (i) the relativistic jet [9], (ii) the disk (a super-Eddington MAD and/or RIAF) [13], (iii) the disk corona, or (iv) the wind/outflow [12].
 }
\label{fig1}
\end{figure}

However, there are some shortcomings in explaining the TDE neutrino by a relativistic jet model. $\sim100$ TDE candidates have so far been observed, of which only three are clearly jetted TDEs [10]. Furthermore, no high-energy gamma-ray and hard X-ray emissions [5] has been observed from AT2019dsg, which would be a clear signature for the production of neutrinos in a relativistic jet. The radio emission of AT2019dsg is too weak to be supportable evidence for the relativistic jet [8]. An off-axis [11] or hidden [12] jet model could however explain some of these inconsistencies.
There are a few alternative models to produce sub-PeV neutrinos from TDEs: an accretion disk, disk corona, and wind/outflow [12,13] (see also Fig.~\ref{fig1}). These are mainly promising for non-jetted TDEs, {for} which the event rate is much higher than the jetted-TDE case. 
For example, the sub-PeV neutrinos are emitted from a super-Eddington magnetically arrested disk (MAD) or a radiatively inefficient accretion flow (RIAF) in the TDE context [13]. Interestingly, the disk's protons accelerate by the second-order Fermi acceleration via disk turbulence, which is different from the relativistic jet model, for which the first-order Fermi acceleration via the shock works. Moreover, high-energy (TeV-scale) gamma-rays are not emitted by efficient pair production in the RIAF model. The main challenge for future multi-messenger studies of TDEs will be to explore whether TDE neutrinos originate from a relativistic jet, accretion disk, disk corona, disk wind/outflow, and other sources. Clues to their production site would come from identifying the acceleration mechanism, cooling processes, hadronuclear and photohadronic interactions, and cascading processes, which differ depending on the given neutrino emitter model [12].

The IceCube-191001A-AT2019dsg association represents the first step in study of high-energy particle emission from TDEs. 
Ongoing and future all-sky-survey telescopes (such as SRG/eROSITA, Vera C. Rubin Observatory Legacy Survey of Space and Time, 
and Einstein Probe) will increase the TDE rate up to the order of thousands per year [10]. Besides, the next-generation IceCube offers a higher sensitivity and a better angular resolution[14]. They will improve the association between astrophysical neutrinos and TDEs. The robust detections of TDE neutrinos will elucidate not only the observed diversity of TDEs but may also help constrain the as-yet-unknown neutrino's mass and lifetime. 
It will be interesting to see if there is any correlation between the electromagnetic radiation variability and the high-energy neutrino emission. 
The IceCube-191001A-AT2019dsg association marks the beginning of multi-messenger observations of TDEs.

\end{document}